\documentclass[a4paper,10pt]{IEEEtran}
\usepackage[utf8]{inputenc}
\usepackage{graphicx}
\usepackage{blindtext}
\usepackage[margin=2cm] {geometry}
\usepackage{setspace}
\usepackage{titling}
\usepackage{caption}
\usepackage{subcaption}
\usepackage{hyperref}
\hypersetup{colorlinks,allcolors=black}
\usepackage{float}
\usepackage{authblk}
\usepackage[acronym,nomain]{glossaries}
\usepackage{tikz}
\usepackage{verbatim}
\usetikzlibrary{arrows,shapes}
\usepackage{mathtools}
\usepackage{blkarray, bigstrut}
\usepackage{algorithm}
\usepackage{algpseudocode}
\usepackage{amsmath}

\RequirePackage[
			style=ieee,
			citestyle=numeric-comp,
			sorting=none,					
			hyperref=true, %
			giveninits=true,
			backend=biber,
			natbib=true,
			url=false,
			isbn=false,
			maxnames=3, 
			maxalphanames=1, 
			doi=false
]{biblatex}	
\algnewcommand\algorithmicforeach{\textbf{for each}}
\algdef{S}[FOR]{ForEach}[1]{\algorithmicforeach\ #1\ \algorithmicdo}

\title{BUbble Flow Field: a Simulation Framework for Evaluating Ultrasound Localization Microscopy Algorithms}
\author[1]{Marcelo Lerendegui}
\author[1]{Kai Riemer}
\author[1]{Bingxue Wang}
\author[2]{Christopher Dunsby}
\author[1]{Meng-Xing Tang}
\affil[1]{Department of Bioengineering, Imperial College London, London, United Kingdom}
\affil[2]{Department of Physics, Imperial College London, London, United Kingdom}
\date{November 2022}

\makeglossaries
\newacronym{am}{AM}{Amplitude Modulation}
\newacronym{bff}{BUFF}{BUbble Flow Field}
\newacronym{cfd}{CFD}{Computational Fluid Dynamics}
\newacronym{cnn}{CNN}{Convolutional Neural Network}
\newacronym{das}{DAS}{Delay and Sum}
\newacronym{lstm}{LSTM}{Long Short Term Memory}
\newacronym{lti}{LTI}{Linear Time Invariant}
\newacronym{mi}{MI}{Mechanical Index}
\newacronym{mb}{MB}{Microbubble}
\newacronym{ode}{ODE}{Ordinary Differential Equation}
\newacronym{pi}{PI}{Pulse Inversion}
\newacronym{psf}{PSF}{Point Spread Function}
\newacronym{rf}{RF}{RadioFrequency}
\newacronym{rnn}{RNN}{Recurrent Neural Network}
\newacronym{ulm}{ULM}{Ultrasound Localization Microscopy}

\newacronym{ius}{IUS}{International Ultrasonics Symposium}
\newacronym{ieee}{IEEE}{Institute of Electrical and Electronics Engineers}

\newacronym{fp}{FP}{False Positive}
\newacronym{tp}{TP}{True Positive}
\newacronym{tn}{TN}{True Negative}
\newacronym{fn}{FN}{False Negative}

\newacronym{hf}{HF}{High Frequency}
\newacronym{lf}{LF}{Low Frequency}

\newacronym{ultra-sr}{ULTRA-SR}{Ultrasound Localisation and TRacking Algorithms for Super Resolution}
\newacronym{epsrc}{EPSRC}{Engineering and Physical Sciences Research Council}

\newacronym{ukri}{UKRI}{UK Research and Innovation}

\begin{document}
\maketitle

\begin{abstract}
Ultrasound contrast enhanced imaging has seen widespread uptake in research and clinical diagnostic imaging. This includes applications such as vector flow imaging, functional ultrasound and super-resolution \gls{ulm}. All of these require testing and validation during development of new algorithms with ground truth data. In this work we present a comprehensive simulation platform \gls{bff} that generates contrast enhanced ultrasound images in vascular tree geometries with realistic flow characteristics and validation algorithms for \gls{ulm}. \gls{bff} allows complex micro-vascular network generation of random and user-defined vascular networks. Blood flow is simulated with a fast \gls{cfd} solver and allows arbitrary input and output positions and custom pressures. The acoustic field simulation is combined with non-linear \gls{mb} dynamics and simulates a range of point spread functions based on user-defined \gls{mb} characteristics. The validation combines both binary and quantitative metrics. \textbf{BFF}'s capacity to generate and validate user-defined networks is demonstrated through its implementation in the \gls{ultra-sr} Challenge at the \gls{ius} 2022 of the \gls{ieee}.
The ability to produce \gls{ulm} images, and the availability of a ground truth in localisation and tracking enables objective and quantitative evaluation of the large number of localisation and tracking algorithms developed in the field. \gls{bff} can also benefit deep learning based methods by automatically generating datasets for training. \gls{bff} is a fully comprehensive simulation platform for testing and validation of novel \gls{ulm} techniques and is open source.
\end{abstract}

\section{Introduction}

Contrast enhanced ultrasound has seen a widespread uptake in research and clinical diagnostic imaging. This includes emerging imaging modalities such as vector flow imaging, super-resolution \gls{ulm} and functional ultrasound. \gls{ulm} is of particular interest as it is the only imaging modality that can visualize vascular structures and flow velocity information at microscopic resolution in deep tissue \textit{in vivo} \cite{dewey_clinical_2020}. This makes ultrasound valuable for clinical applications where tissue microvascular flow is of interest. For example, cancer research has shown that increased angiogenesis is an early event in the development of tumours \cite{laitakari_size_2004}. In the case of atherosclerosis, which can lead to strokes and cardiac arrest, angiogenesis is the predominant form of neovascularization \cite{noauthor_neovascularization_nodate} and \gls{ulm} could have the potential to be used for screening. 

Most ultrasound contrast agents are highly compressible gas \glspl{mb} surrounded by a lipid monolayer. \glspl{mb} behave as scatterers due to the difference in acoustic impedance between the gas and the surrounding blood. But \glspl{mb} also respond non-linearly to the incident pressure wave due to their compressible gas component. Taking advantage of their non-linear response, contrast sequences such as \gls{pi} or \gls{am} allow their easy separation from regular tissue. \gls{ulm} localizes isolated \glspl{mb} and tracks them over many frames as they flow through the vascular network. Through the localization \gls{ulm} creates sub wavelength resolution vessel representations and flow velocity maps. 

The requirement for isolated bubbles poses a sparsity constraint that generates a series of limitations on the \gls{mb} concentration that can be used and the total acquisition time that is needed. Improving the performance of \gls{ulm} towards the goal of real-time \gls{ulm} is desired. This puts the focus on providing better bubble separation and increasing the localization precision. Efforts have been made to increase the speed of \gls{ulm} for example by localizing \glspl{mb} at higher densities through deconvolution and multi-feature tracking \cite{yan_super-resolution_2022}, through sparsity-based methods \cite{bar-zion_sushi_2018}, by splitting the bubble signal in the 3D Fourier domain \cite{huang_short_2020} or through the use of phase change contrast agents \cite{riemer_fast_2022}. However, to evaluate the large number of localisation and tracking algorithms properly, realistic datasets with ground truth are required. For \gls{ulm}, such datasets should comprise realistic \gls{mb} \glspl{psf} that coherently interact when overlapped, and the flow should reflect the structures and behavior of the clinical end-application. Datasets also need to be large and contain tens of thousands of \glspl{mb} to aid in the design and network training of the increasingly popular deep learning based \gls{ulm} methods \cite{van_sloun_super-resolution_2021}.  

There are many tools for simulating the propagation of ultrasound wave fields. Two prominent examples are Field II \cite{jensen_field_1996} and k-Wave \cite{treeby_k-wave_2010}. There are also equations that model the behavior of \glspl{mb}, and commercial tools for computing flow such as StarCCM \cite{siemens_digital_industries_software_simcenter_2021}. But they are not specifically designed for \gls{ulm} and they are difficult to combine. Random bivariate Gaussian curves have been used as \glspl{psf} for training \gls{cnn} models \cite{liu_deep_2020, van_sloun_super-resolution_2021}. But these are limited due to the lack of sidelobes or bubble tails, and the interaction between overlapping \glspl{psf} is not coherent. Womersley flow has been used to perform fast simulation of flow on large vessels \cite{riemer_contrast_2022}. But this method yields results for a single cylindrical section of a vessel, it is not a realistic model for microvasculature, and is computationally expensive to solve. There have been several attempts to model the fluctuation of a \gls{mb}'s radius when exposed to an incident pressure field. The most basic form is the Rayleigh-Plesset equation \cite{versluis_ultrasound_2020} that models a gas bubble in an infinite pool of liquid, with the limitation of its assumption that the bubble boundary moves slower than the speed of sound. Further models do consider the acoustic emission of the bubble \cite{versluis_ultrasound_2020}, but not the behavior of the lipid monolayer. ULM normally insonifies \glspl{mb} with pressure high enough to produce significant non-linear response. A more recent model added shell buckling behavior with compression and rupture on large oscillations, which considers the physical properties of the lipid monolayer and its possible buckled, elastic and broken states \cite{marmottant_model_2005}. A more general approach was demonstrated by combining k-Wave and the Marmottant model \cite{brown_investigation_2019}. This method included reflections and aberrations produced by the non-linear propagation on the tissue, but it is highly demanding on computing power and memory, and requires a spatial grid to be defined which affects the accuracy of the \gls{mb} locations. None of the efforts so far are designed for large dataset generation, and comprise all four essential components: bubble response, acoustic fields, flow behavior and validation. 

To address the shortcomings we created a fully comprehensive simulation platform for \gls{ulm}, called \gls{bff}, that incorporates arbitrary micro-vessel network generation with a quick \gls{cfd} solver and linear acoustic propagation with non-linear \gls{mb} dynamics and subsequent binary and quantitative evaluation for \gls{ulm} algorithms. In the following we will describe how \gls{bff} uses a custom tool to generate organic microvessel structures, how the Hagen–Poiseuille model for quick \gls{cfd} simulation is calculated, how Field II as the acoustic propagation backend is combined with the Marmottant \gls{ode} to simulate the response of the \glspl{mb} and how the evaluation of \gls{ulm} is achieved by means of the implementation of \gls{bff} in the Ultrasound Localisation and TRacking Algorithms for Super Resolution (ULTRA-SR) Challenge at the \gls{ius} 2022 of the \gls{ieee}.

\section{Methods}

\subsection{Microvascular Flow}
To model microvascular structures we first need to briefly define microvascular flow. The Reynolds number describes the ratio of inertial forces to viscous forces. In small vessels of the microvasculature the Reynolds number is very low and radial components of flow are zero considering that flow is laminar for $Re < 2300$  

\begin{equation}\label{eqn:reynolds}
Re = \frac{u D}{\mu}
\end{equation}

Where $u$ is the flow velocity, $D$ the vessel diameter and $\mu$ the kinematic viscosity of blood. Furthermore, such flow is quasi steady, fully developed and vessels segments are assumed to be rigid and straight. Subsequently, the axial momentum equation of the flow field derived from the Navier Stokes equation for incompressible Newtonian fluids reduces to
 
\begin{equation} \label{eqn:axial_momentum}
\frac{1}{r}\frac{\delta}{\delta r}\left( r \frac{\delta u_z}{\delta r} \right) = - \frac{1}{\mu}\frac{\delta p}{\delta z}
\end{equation}

where $\delta p/\delta z$ is the pressure as a function of the axial coordinate $z$ and $r$ is the radial coordinate. Imposing a no-slip condition on the wall ($u=0$ at $r=R$) a parabolic shape of the velocity profile can be determined

\begin{equation} \label{eqn:parabolic_profile}
u_z(r) = u_{z,max} \left( 1 - \frac{r^{2}}{R^{2}}\right)
\end{equation}

where $u_{z,max}$ is the maximum velocity, which is twice the mean velocity ($u_{z,mean} = 1/2 u_{z,max}$) for a parabolic (cross-sectional) flow profile   

\begin{equation}\label{eqn:max_velocity}
u_{z, max} = \frac{R^{2}}{4\mu}\frac{\Delta p}{l} 
\end{equation}

The volume flow in a vessel with finite length $l$ can be calculated with the Hagen–Poiseuille law using the hydrodynamic resistance $\xi$ and the pressure difference $\Delta p$.

\begin{equation} \label{eqn:hagen_poiseuile}
Q = \frac{\pi r^{4}}{8\mu l} \Delta p = \frac{\Delta p}{\xi}
\end{equation}

These basic equations of flow and pressure will be used in the network solver as a system of linear equations.

\subsection{Network Generation}
\gls{bff} implements an organic vessel tree generation tool based broadly in a recurrent process of vasculogenesis and sprouting angiogenesis. Vessel structures are approximated as weighted directed graphs with nodes and edges. Each edge represents a cylindrical section of a vessel, with characteristics of radius, length and orientation; and each node represents the connection between two edges. The final output of this tool is a randomized binary tree structures that can be fully customized to mimic what is seen \textit{in vivo}.

The process of network creation starts with an initial state defined by: initial position, orientation, radius, and recursion level. A vessel is created by iteratively appending new edges. Each new edge will have new randomized properties based on the current state: the rotation both in elevation and asimuth can be tuned to different vessel tortuosities, and the radius can be changed as new segements are added, creating a radius decay along the vessel. Random locations along the created vessel are chosen as 'sprouts' or locations for bifurcations, and the process is run recursively, creating new vessels at those locations. The whole process is highly customizable. This is achieved by defining parameters as functions of the current status of the generation algorithm: $param=f(n, d, e1, e2, r, lvl)$ with:

\begin{itemize}
    \item $n$: the current node
    \item $d, e1, e2$: the current edge orientation (orthonormal reference system)
    \item $r$: the current radius
    \item $lvl$: the current recursion level    
\end{itemize}

The network generation can be constraint by a global set of parameter as shown in Table \ref{tbl:net_gen_params}. Organs usually have specific shapes, and that shape limits the extent of its vessel structure. For example the $inside\_f$ function is used for creating specific organ shapes, by limiting the generation of vessels only to positions inside the shape. 

\begin{figure}
\begin{table}[H]

    \begin{tabular}{ p{2.5cm}|p{5cm}}
            \textbf{Parameter} & \textbf{Description}
        \\\hline
            \textbf{edge{\_}step{\_}f} & step size in meters
        \\\hline
            \textbf{inside{\_}f} & when to stop generating
        \\\hline
            \textbf{rot{\_}f} & new orientation to use on next section of current vessel
        \\\hline
            \textbf{r{\_}decay{\_}f} & new radius to use on next section of current vessel
        \\\hline
            \textbf{bif{\_}occurs{\_}f} & whether or not a bifurcation occurs at this place
        \\\hline
            \textbf{bif{\_}r{\_}decay{\_}f} & new radius to use on new branch vessel
        \\\hline
            \textbf{bif{\_}rot{\_}f} & new orientation to use on new branch vessel
    \end{tabular}
    
    \captionsetup{justification=justified}
    \caption{The global network generation parameters can be used to constrain the network generation. For example, edge{\_}step{\_}f describes the spatial frequency of edges and is inside{\_}f determines the outlining 3D shape of the network.} 
    \label{tbl:net_gen_params}
    
\end{table}
\end{figure}

\subsection{Network Solver}
The flow and pressure in a vessel network can be solved for every position as a system of linear equations. To construct the equations, the node-edge incidence matrix, input and output pressures, and the hydrodynamic conductance of all the edges are required. In addition, the following assumptions are made: 
\begin{itemize}
    \item fluid is incompressible
    \item inputs and outputs are hanging nodes 
    \item a hanging node is connected to only one edge
\end{itemize}

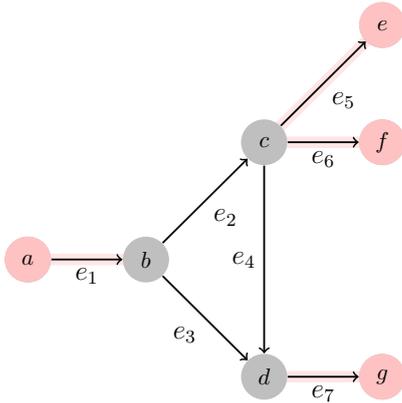
\begin{figure}[ht]
\centering
\resizebox {0.7\columnwidth} {!} {

\tikzset{
  vertex/.style={circle, fill=black!25, minimum size=20pt},
  highlighted_vertex/.style={vertex, fill=red!24},
  edge/.style={draw, thick, ->},
  highlighted_edge/.style={draw, line width=5pt, -, red!10},
  weight/.style={font=\large},
}

\pgfdeclarelayer{background}
\pgfsetlayers{background, main}

\begin{tikzpicture}[scale=1.8, auto,swap]
    \foreach \pos/\name in {
        {(0,2)/a},
        {(1,2)/b},
        {(2,3)/c},
        {(2,1)/d},
        {(3,4)/e},
        {(3,3)/f},
        {(3,1)/g}}
    \node[vertex] (\name) at \pos {$\name$};
    
    \foreach \source/ \dest /\weight in {
        a/b/e_1,
        b/c/e_2,
        b/d/e_3,
        c/d/e_4,
        c/e/e_5,
        c/f/e_6,
        d/g/e_7}
    \path[edge] (\source) -> node[weight] {$\weight$} (\dest);

    \foreach \vertex in {
        a,
        e,
        f,
        g}
    \path node[highlighted_vertex] at (\vertex) {$\vertex$};
    
    \begin{pgfonlayer}{background}
        \foreach \source / \dest in {
            a/b,
            c/e,
            c/f,
            d/g}
        \path[highlighted_edge] (\source.center) -- (\dest.center);
    \end{pgfonlayer}
\end{tikzpicture}
}
\caption{Simple sample network to illustrate the relation between node (a-f) and edges (1-7). Nodes connected to only one edge are hanging nodes. Their corresponding edge is highlighted.}
\label{fig:sample_network}
\end{figure}

Vessel networks are represented as a weighted directed graph, containing nodes and edges illustrate in Figure \ref{fig:sample_network}.

The incidence matrix \textit{I} encodes the structure of the vessel network. The matrix \textit{I} can be used to express a system of linear equations that relate the edge pressure differences and the node pressures, and another system of linear equations that relates the current at each node. The edge-node incidence matrix for the sample network is

\begin{equation*}
\mathbf{I}=
\begin{blockarray}{r ccccccc}
    \begin{block}{r ccccccc}
      & a & b & c & d & e & f & g \\
    \end{block}
    \begin{block}{r [ccccccc]}
        e_1 &  1  & -1  &  0  &  0  &  0  &  0  &  0 \\
        e_2 &  0  &  1  & -1  &  0  &  0  &  0  &  0 \\
        e_3 &  0  &  1  &  0  & -1  &  0  &  0  &  0 \\
        e_4 &  0  &  0  &  1  & -1  &  0  &  0  &  0 \\
        e_5 &  0  &  0  &  1  &  0  & -1  &  0  &  0 \\
        e_6 &  0  &  0  &  1  &  0  &  0  & -1  &  0 \\
        e_7 &  0  &  0  &  0  &  1  &  0  &  0  & -1 \\
    \end{block}
\end{blockarray}
\end{equation*}

The pressure at the hanging nodes is known from the start, but the pressure at non-hanging nodes are unknowns that need to be solved. The incidence matrix is separated into two: $I_h$ and $I_{nh}$, containing the columns of $I$ corresponding to hanging nodes and non-hanging nodes: 

\begin{equation*}
\mathbf{I_h}=
\begin{blockarray}{r cccc}
    \begin{block}{r cccc}
      & a & e & f & g \\
    \end{block}
    \begin{block}{r [cccc]}
        e_1 &  1  &  0  &  0  &  0  \\
        e_2 &  0  &  0  &  0  &  0  \\
        e_3 &  0  &  0  &  0  &  0  \\
        e_4 &  0  &  0  &  0  &  0  \\
        e_5 &  0  &  1  &  0  &  0  \\
        e_6 &  0  &  0  &  1  &  0  \\
        e_7 &  0  &  0  &  0  &  1  \\
    \end{block}
\end{blockarray}
,
\mathbf{I_{nh}}=
\begin{blockarray}{r ccc}
    \begin{block}{r ccc}
      & b & c & d \\
    \end{block}
    \begin{block}{r [ccc]}
        e_1 & -1  &  0  &  0  \\
        e_2 &  1  & -1  &  0  \\
        e_3 &  1  &  0  & -1  \\
        e_4 &  0  &  1  & -1  \\
        e_5 &  0  &  1  &  0  \\
        e_6 &  0  &  1  &  0  \\
        e_7 &  0  &  0  &  1  \\
    \end{block}
\end{blockarray}
\end{equation*}

The pressure difference across edges can be expressed as:
\begin{equation} \label{eqn:edge_node_pressure}
    P_e = I_h P_0 - I_{nh} P_n
\end{equation}
with $P_0$ the known pressure at each hanging node, and $P_n$ the unknown pressure at non-hanging nodes. Given the assumption that the fluid is incompressible, the sum of flows in each non-hanging node must be zero. This can be expressed as a system of linear equations:

\begin{equation} \label{eqn:flow_node}
I_{nh}^{T} Q_e = 0
\end{equation}

Equation \autoref{eqn:hagen_poiseuile} relates the pressure difference at the ends of an edge to the volume flow going through it. This can be expanded into a system of equations for each edge as:

\begin{equation} \label{eqn:edge_pressure_flow}
Q_e = C P_e
\end{equation}
with $P_e$, edge pressure difference vector; $Q_e$, Edge volumetric flow vector; and $C$, Edge Flow conductance Matrix

The edge conductance matrix is a diagonal matrix whose values be obtained by calculating the flow conductance at each edge, using its radius and the viscosity of the liquid on \autoref{eqn:hagen_poiseuile}:

\begin{equation*}
\mathbf{C}=
\begin{blockarray}{r ccccccc}
    \begin{block}{r ccccccc}
      & e_1 & e_2 & e_3 & e_4 & e_5 & e_6 & e_7 \\
    \end{block}
    \begin{block}{r [ccccccc]}
        e_1 &   \frac{1}{\xi_1}  &  0  &  0  &  0  &  0  &  0  &  0 \\
        e_2 &  0  &   \frac{1}{\xi_2}  &  0  &  0  &  0  &  0  &  0 \\
        e_3 &  0  &  0  &   \frac{1}{\xi_3}  &  0  &  0  &  0  &  0 \\
        e_4 &  0  &  0  &  0  &   \frac{1}{\xi_4}  &  0  &  0  &  0 \\
        e_5 &  0  &  0  &  0  &  0  &   \frac{1}{\xi_5}  &  0  &  0 \\
        e_6 &  0  &  0  &  0  &  0  &  0  &   \frac{1}{\xi_6}  &  0 \\
        e_7 &  0  &  0  &  0  &  0  &  0  &  0  &   \frac{1}{\xi_7} \\
    \end{block}
\end{blockarray}
\end{equation*}

Given \autoref{eqn:edge_node_pressure}, \autoref{eqn:flow_node}, and \autoref{eqn:edge_pressure_flow} we can replace \autoref{eqn:edge_pressure_flow} in \autoref{eqn:flow_node} to obtain:

\begin{equation}
I_{nh}^{T} C P_e = 0
\end{equation}

by replacing \autoref{eqn:edge_node_pressure} we get:

\begin{equation}
I_{nh}^{T} C  (I_h P_0 - I_{nh} P_n) = 0
\end{equation}

\begin{equation}
I_{nh}^{T} C I_h P_0 - I_{nh}^{T} C I_{nh} P_n = 0
\end{equation}

\begin{equation}
I_{nh}^{T} C I_{nh} P_n = I_{nh}^{T} C I_h P_0
\end{equation}

which can be expressed as a linear system of equations in the form of $M x = b$ with:
\begin{itemize}
    \item $M = I_{nh}^{T} C I_{nh}$
    \item $x = P_n$
    \item $b = I_{nh}^{T} C I_h P_0$
\end{itemize}

This will solve for the pressure at each non-hanging node, and, using equations \autoref{eqn:edge_node_pressure}, \autoref{eqn:edge_pressure_flow}, the volumetric flow at each edge can be derived. Note, a hanging node can be placed anywhere within the network.

\subsection{Track generation}
The bubble positions and their tracks through the network implemented through a fast track generation algorithm based three conditions:
\begin{itemize}
    \item the probability of a bubble taking a path at a bifurcation is proportional to the flow in that path.
    \item bubbles move only in a streamline (they maintain their relative radial position through the whole network according to the rules of laminar flow).
    \item the number of bubbles is much greater than the number of all possible tracks
\end{itemize}

All possible tracks or root-to-leaf paths can be extracted from the network by computing all the combinations of paths at each vessel bifurcation. At each bifurcation, a bubble can randomly go through one of the paths. The probability of choosing one is calculated based on flow conservation and fluid incompressibility. The volume flow rate of the inlet $Q_1$ is the sum of its outlet $Q_2$ and $Q_3$: 
\begin{equation}
    Q_1 = Q_2 + Q_3    
\end{equation}

The fraction of the incoming volume that will go to branch two will then be $\frac{Q_2}{Q_1}$, so we can approximate the probability of taking that branch as:

\begin{equation}
    p = \frac{Q_2}{Q_1}
\end{equation}

The probability of a bubble traversing a whole track is calculated as the product of all bifurcation probabilities in a track:

\begin{equation} \label{eqn:trac_probability}
p_{track} = \prod_{i \in T}^{} \frac{edge_i.Q}{edge_{i-1}.Q} 
\end{equation}

To generate the ground truth information, any number of bubbles are moved with the network flow. For each bubble, a track is randomly chosen using the track probability calculated with \autoref{eqn:trac_probability}. A starting position is chosen for the bubbles comprising both radial, angular and axial position in the starting vessel. From the starting position, a particle will move through edges at a constant speed of $edge.vel * ( 1-r^2 )$, with $edge.vel$ the (max) velocity at the center of the vessel, and $r$ is the fractional radius of the streamtube the particle is traveling in. The particle position is updated using its velocity and the $\Delta t$. Once the particle reaches the end of the current edge, the overshoot is corrected and the starting position for the next edge defined:
\begin{equation} \label{eqn:overshoot}
pos = next\_edge.start + overshoot * \frac{next\_edge.vel}{cur\_edge.vel}
\end{equation}
where $next\_edge.start$ is the starting location of the next edge, $next\_edge.vel$ is its velocity and $cur\_edge.vel$ the current edge velocity. This process is repeated until the bubble reaches the end of the track. The result of this simulation is a table of events with in the format $[frame, bubble_id, x, y, z]$. Additional parameters such as particle velocity or tube radius can be added to the table as ground truth if desired. For the final output table, all the tables from different bubbles are stacked vertically and sorted by frame number.

\subsection{Non-linear Bubble Simulation}
\gls{bff} can solve for any \gls{ode} to obtain the response of a \gls{mb}, as a default, we chose to use a modified Rayleigh–Plesset equation as follows \cite{marmottant_model_2005}:

\begin{multline}  \label{eqn:marmottant_ode}
    \rho_l (R \ddot{R}+ \frac{3}{2}\dot{R}^2) = [ P_0 + \frac{2 \sigma (R_0)}{R_0} ](\frac{R}{R_0})^{-3\kappa} (1-\frac{3 \kappa}{c} \dot{R}) \\
    - P_0 - \frac{2\sigma(R))}{R} - \frac{4\mu\dot{R}}{R} - \frac{4\kappa_s\dot{R}}{R^2} - P_{ac}(t)
\end{multline}

\begin{equation} \label{eqn:marmottant_sgm}
\sigma(R) = 
     \begin{cases}
       0 & \text{if} R \leq R_{buckling} \\
       \chi (\frac{R^2}{R_{buckling}^2} - 1) & \text{if} R_{buckling} \leq R \leq R_{break} \\
       \sigma_{water} & \text{if ruptured and } R \geq R_{ruptured}       \\
     \end{cases}
\end{equation}

This \gls{ode} is solved using Runge-Kutta methods at a higher sampling rate than the ultrasound simulation, with an user-defined oversampling factor. Solving the \gls{ode} yields the bubble radius over time and its first two derivatives: $R$, $\dot{R}$, and $\ddot{R}$. The scattered pressure at a distance $d$ from the center of the bubble is defined \cite{vokurka_rayleighs_1985}:

\begin{equation} \label{eqn:bub_scatered}
P=\frac{\rho_l}{d}(R^2 \ddot{R} + 2R\dot{R}^2)
\end{equation}

\gls{bff} has implementations of both a custom bubble that users can setup, and a commercial agent such as SonoVue\textsuperscript{\tiny\textregistered} (sulphur hexafluoride) that is pre-configured with fitted parameters \cite{marmottant_model_2005} as shown in Table \ref{tbl:mb_params}).

\begin{table}[H]
    \begin{center}
    \captionsetup{justification=centering}
    \begin{tabular}{ |c|p{.5\columnwidth}|c| }
        \hline
            \textbf{Param} & \textbf{Description} & \textbf{SonoVue}
        \\\hline
            $\rho_l$ & Density of surrounding liquid & $10^3 \frac{Kg}{m^3}$
        \\\hline
            $\sigma_l$ & Surface tension of surrounding liquid & $0.073 \frac{N}{m}$
        \\\hline
            $\mu_l$ & Viscosity of surrounding liquid & $2.0 10^{-3} P s$
        \\\hline
            $\kappa$ & Polytropic gas exponent & $1.095$
        \\\hline
            $\kappa_s$ & Surface dilatational viscosity from the mono-layer & $ 7.2 10^{-9} N$
        \\\hline
            $\chi$ & Elastic compression modulus of the mono-layer & $1.0 \frac{N}{m}$
        \\\hline
            $R_0$ & Equilibrium radius of the bubble & $0.975 \mu m$
        \\\hline
            $R_{buckle}$ & Lower radius limit of the elastic state & $0.975 \mu m$
        \\\hline
    \end{tabular}
    
    \caption{Microbubble Simulation Parameters for a bubble of radius $3\mu m$} \label{tbl:mb_params}
    
    \end{center}
\end{table}

\subsection{Integration into Linear Acoustic Simulation}
Field II is a platform that simulates pressure fields from arbitrarily shaped transducers \cite{jensen_field_1996}.
It models the ultrasound system as a Linear Time Invariant system which is characterized by it's impulse response; the response to a Dirac excitation. This assumption of linearity makes Field II fast, allowing it to divide the problem into parallelizable parts, by splitting each transducer element into small mathematical sub-elements, and making every scatterer independent. Over the years, Field II's results have been shown to be accurate with ultrasound experiments \cite{jensen_calculation_1992}. The assumption of linearity is broken when incorporating the non linear dynamics of the contrast agent. To integrate bubble simulation into the FieldII's acoustic simulation, a multi step process is required. It is assumed that bubbles are independent and there is no coupling between them.

First, the transmitted pressure signal at the location of the bubble is calculated. This is done by simulating the transducer and calculating the pressure over time at the bubble location. Second, using the pressure signal calculated at step 1 as an input, we calculate the bubble response signal by using equation \autoref{eqn:marmottant_ode} and then equation \autoref{eqn:bub_scatered}. Finally, the bubble response signal is propagated back to the transducer. Field II does not have a direct way to calculate the signal received with the transducer when transmitting from a point source in space. This can be done by convolving the bubble response and the spatial impulse response from the bubble position to the transducer, which, because of the reversibility property of ultrasound, is equal to the impulse response from the transducer to a \gls{mb} position.

\begin{equation}
    RX(t) = \text{bub{\_}resp}(t) \ast H(t, \text{bub{\_}pos})
\end{equation}

\subsection{Evaluation} 
The evaluation metrics of \gls{bff} present estimations of both the fractions of correct events and the error for localization and tracking of an \gls{ulm} algorithm. 

\paragraph{Localization Evaluation}
A localization is considered a \gls{tp} if it lies close enough to a ground truth position; the localisation is inside a defined search radius centered around the ground truth location. Following this logic, a ground truth \gls{mb} with no localisation in its search radius is considered a \gls{fn}, and a localization that is not inside the search radius of any ground truth \gls{fp} location is a \gls{fp}. Two metrics, \textit{precision} and \textit{recall} are used to indicate the fractions of the correct localisation:

\begin{equation} \label{eqn:loc_precision}
Precision = \frac{TP}{TP + FP}
\end{equation}

\begin{equation} \label{eqn:loc_recall}
Recall = \frac{TP}{TP + FN}
\end{equation}

A third metric which describes the average distance between the true position of a \gls{mb} (B) and the localisation position (L) is calculated to reflect the root mean square error (RMSE) of all true positive localisation ($N_{TP}$):

\begin{equation} \label{eqn:track_mean_err}
RMSE = \frac{\sum_{i\in TP} dist(B_i \leftrightarrow L_i)}{N_{TP}}
\end{equation}

\paragraph{Tracking Evaluation}

Tracking is a process that analyzes localizations over consecutive frames and determines pairs of localizations that correspond to the same bubble. Tracking is strongly dependent on the localization performance. A bubble that was not localized will never be paired, and a \gls{fp} localization should never be paired. If one were to consider all localizations for tracking, \gls{fn} and \gls{fp} localizations would generate a double penalty on the erroneous localizations. To separate the tracking evaluation from the localization evaluation, only true positive localizations must be considered, i.e. only pairings of correctly localized bubbles in both frames are analyzed. This diminishes the total number of pairings to work with. As the total number of localisation increases the impact of this systematic bias is reduced. 

Across two consecutive frames, two paired localizations that have the same localization id are considered \gls{tp} if their associated ground truth bubbles have the same bubble id. Any bubble pair that doesn't have an associated localization pair is considered a \gls{fn}. Any localization pair that is not associated with a bubble pair is considered a \gls{fp}. Tracking is evaluated both in fraction of correct pairs and in effective pair distance. Precision and Recall are used to indicate the fractions of correct pairings:

\begin{equation} \label{eqn:track_precision}
Precision = \frac{TP}{TP + FP}
\end{equation}

\begin{equation} \label{eqn:track_recall}
Recall  = \frac{TP}{TP + FN}
\end{equation}

There is no direct way of measuring distance error for a pairing process.
The correct pair distance can be used, i.e. the distance a bubble travels from one frame to the other.
It is a fair metric, bubbles that travel further away are more difficult to track, which would yield a better result in the metric.
Given that the total number of pairs is dependent on the localization process, the fraction of correct paired distance. This is equivalent to a weighted Jaccard index with the weights being the distance of each pair.

\begin{equation} \label{eqn:track_jaccaard}
J = \frac{TP_d}{TP_d + FP_d + FN_d}
\end{equation}
where:
\begin{equation}
TP_d = \sum_{i\in TP} dist(pair_i) 
\end{equation}

\begin{equation}
FP_d = \sum_{i\in FP} dist(pair_i) 
\end{equation}

\begin{equation}
FN_d = \sum_{i\in FN} dist(pair_i) 
\end{equation}

To provide better clarity, a remapping of this metric of the form $L=2*J-1$ is used, so the range of the metric goes from $[0,1] \rightarrow [-1, 1]$. A negative number means track is mostly wrong, and positive number is mostly right. The final metric has then the following expression:
\begin{equation} \label{eqn:track_metric}
J_{map} = \frac{TP_d - FP_d - FN_d}{TP_d + FP_d + FN_d}
\end{equation}

\subsection{BFF Validation Experiments} 
\subsubsection{Network Generation}
Multiple networks were generated with random parameters. In addition, a kidney and liver were mimicked to illustrate the ability of generating organic like structures with little effort. The network generator was manually constrained only by the $inside\_f$ to yield a similar geometric shape.

\subsubsection{Bubble Simulation}
A simulation was performed to validate \gls{bff}'s non linear simulation capabilities and replicated \textit{in vitro}. The purpose was to test for \glspl{psf} realism and variety. The \textit{in-vitro} experiment was done with an L11-5v 128-element linear array transducer with a tansmitted frequency of 7 MHz. Perfluorobutane \glspl{mb} where created following the process described in \cite{hansen-shearer_ultrafast_2022}, and placed in a beaker with de-gassed, filtered water at room temperature. Images were acquired with a Verasonics Vantage 256 platform (Verasonics Inc., Redmond, WA). Plane wave imaging compounding using three angles and a transmission amplitude corresponding to a \gls{mi} of 0.05 was used. The simulation replicated the transducer, and \glspl{mb} with random parameters and locations. The resulting images where visually assessed and \glspl{psf} of similar morphology were identified and extracted.

\subsubsection{\gls{ultra-sr} Challenge}
\gls{bff}'s qualities were demonstrated at the \gls{ultra-sr} challenge at the \gls{ius} 2022 of the \gls{ieee}. Contrast enhanced ultrasound videos from four randomly generated networks with ground truth were produced for this competition. The proposed evaluation criteria were used for objective and quantitative evaluation of the participants localisation and tracking algorithms. The synthetic data was created using networks with varying density on different spatial sections. To minimize the predictability of the path of a \gls{mb} the vessel tree generation parameters e.g., bifurcation probability, angle between consecutive segments or \gls{mb} seeding probability were randomized. A training dataset was also provided with a single branch with a low \gls{mb} concentration. The competition data combined multiple branching vessel trees into a single network with high \gls{mb} concentration.

\section{Results}

\subsection{Network Generation}
\begin{figure*}[ht!]
  \centering
    \includegraphics[width=1\columnwidth*2]{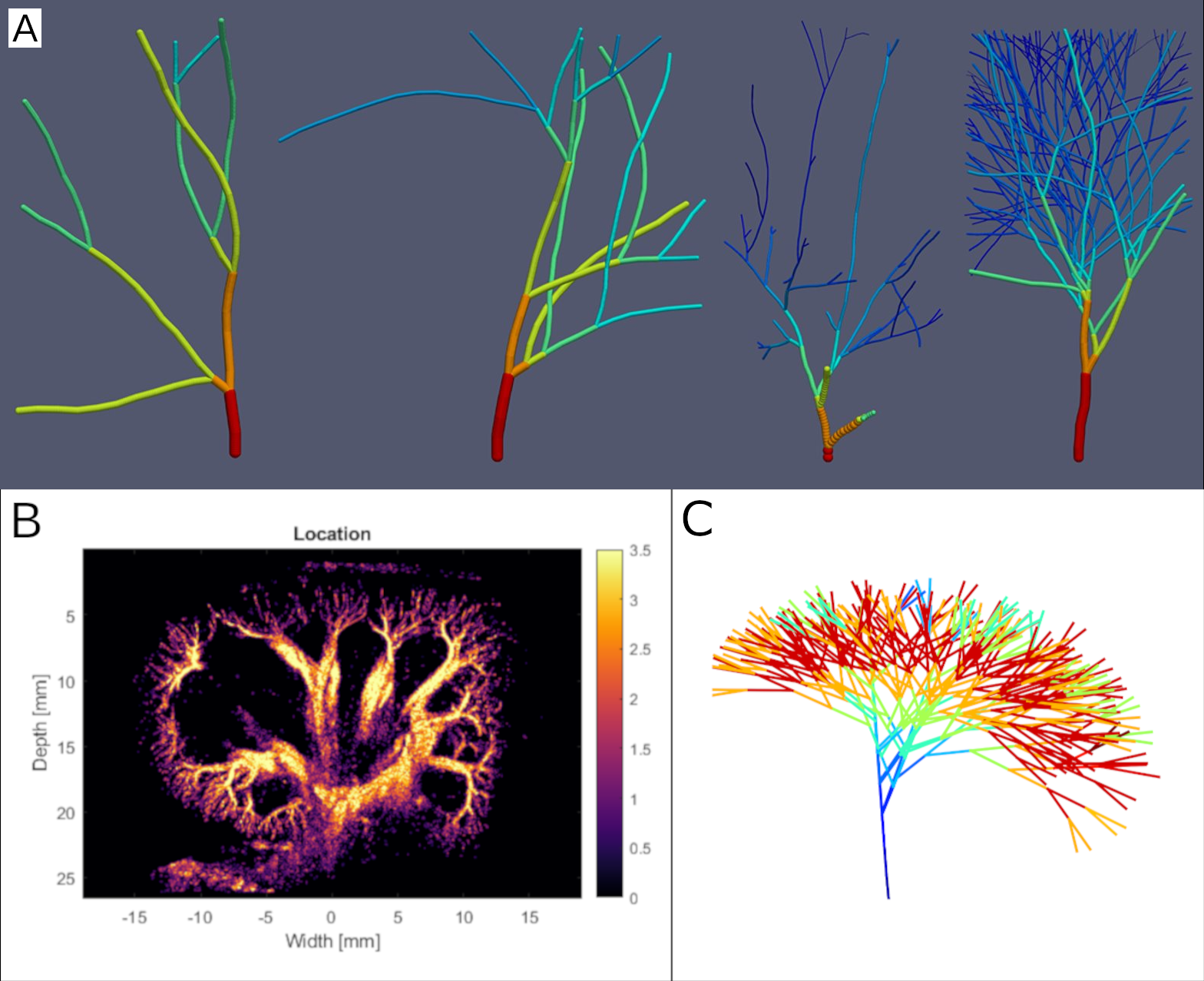}
      \caption{\textbf{A:} Randomly generated networks obtained using different maximum allowed recursion levels. Recursion level from left to right: lvl=2, lvl=3, lvl=4, lvl=5. Rectangular constraint on network shape. \textbf{B:} Superresolution Ultrasound Image of a Rabbit Kidney\cite{riemer_fast_2022}. \textbf{C:} Generated Network mimicking a the ramifications of one of the interlobar arteries of the rabbit's kidney.}
  \label{fig:custom_nets}
\end{figure*}

Figure \ref{fig:custom_nets} (A) shows randomly generated networks with a different number of maximum allowed recursion levels. The vessels are color coded according to their size. Different amount of tortuosity and three-dimensionality are demonstrated with complexity of the network increasing from left to right. Networks are constrained e.g. by the $inside\_f$ function to a rectangular shape. Figure \ref{fig:custom_nets} (B, C) shows the ability of \gls{bff} to generate physiological shapes based on such simple constraints. \gls{bff} is used to generate a network closely resembling the corony microvascularture (B) and the shape of a kidney (C). The parameterised generation of networks can generate any arbitrary shape.    

\subsection{Network Seeding}
\begin{figure}[H]
  \centering
    \includegraphics[width=1\columnwidth]{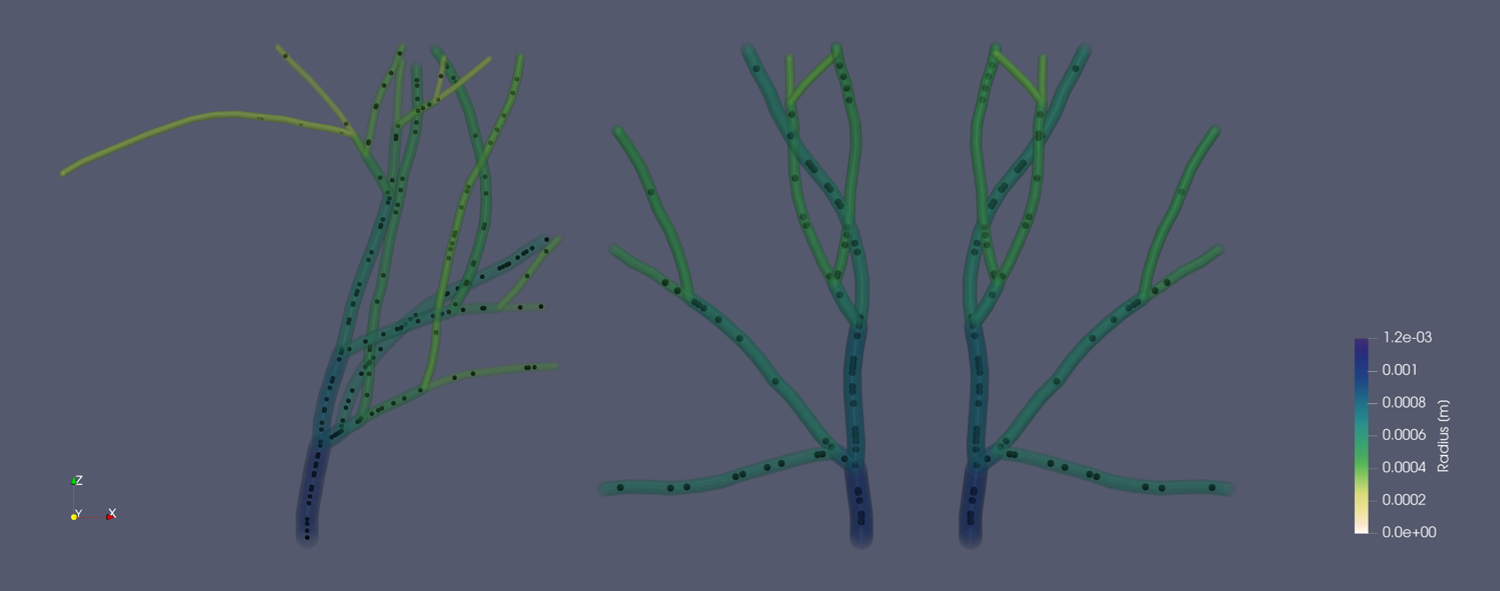}
    \caption{Particles seedded on the network}
    \label{fig:net_seeding}
\end{figure}

\autoref{fig:net_seeding} shows the seeding of a randomly generated network along the tracks with a bubble radial position r=0.

\subsection{Bubble Simulation}
\begin{figure}[H]
  \centering
    \includegraphics[width=0.8\columnwidth]{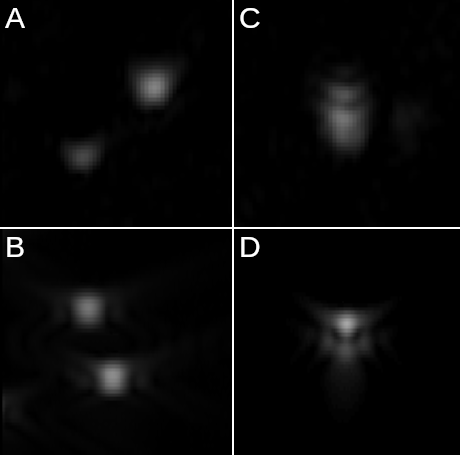}
    \caption{Comparison between \gls{psf} morphology seen in simulation and \textit{in vitro}. \textbf{A \& C:} Mono-lobe and multi-lobe morphologies observed at \textit{in vitro} beaker experiments. \textbf{B \& D:} Mono-lobe and multi-lobe morphologies can also be observed when recreating the experiment using \gls{bff}.}
    \label{fig:psf_compare}
\end{figure}

\gls{bff}'s parameterised bubble generation results in a variety of shapes of the point spread function associated with a \gls{mb}. Figure \ref{fig:psf_compare} qualitatively shows the difference in the B-Mode image between simulated \glspl{mb} (B, D) and \glspl{mb} from an \textit{in vitro} beaker acquisition (A, C). The mono-lobe (A, B) and multi-lobe (C, D) morphological shapes are distinct from each other. The side lobe signal is stronger in the simulation compared to the \textit{in vitro} experiment.   

\subsection{\gls{ultra-sr} Challenge}

\begin{figure}[H]
  \centering
    \includegraphics[width=1\columnwidth]{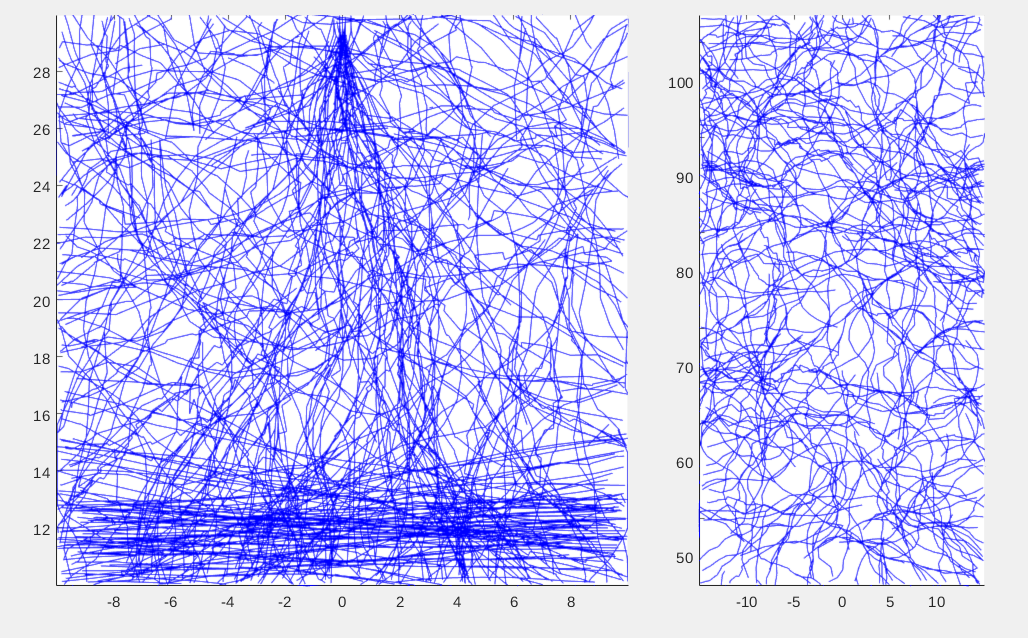}
      \caption{Structures generated with \gls{bff} for the \gls{ultra-sr}. \textbf{A:} The vessel network for the \gls{hf} simulation. \textbf{B:} The vessel network for the \gls{lf} simulation}
  \label{fig:net_struct}
\end{figure}

\begin{figure}[H]
  \centering
    \includegraphics[width=1\columnwidth]{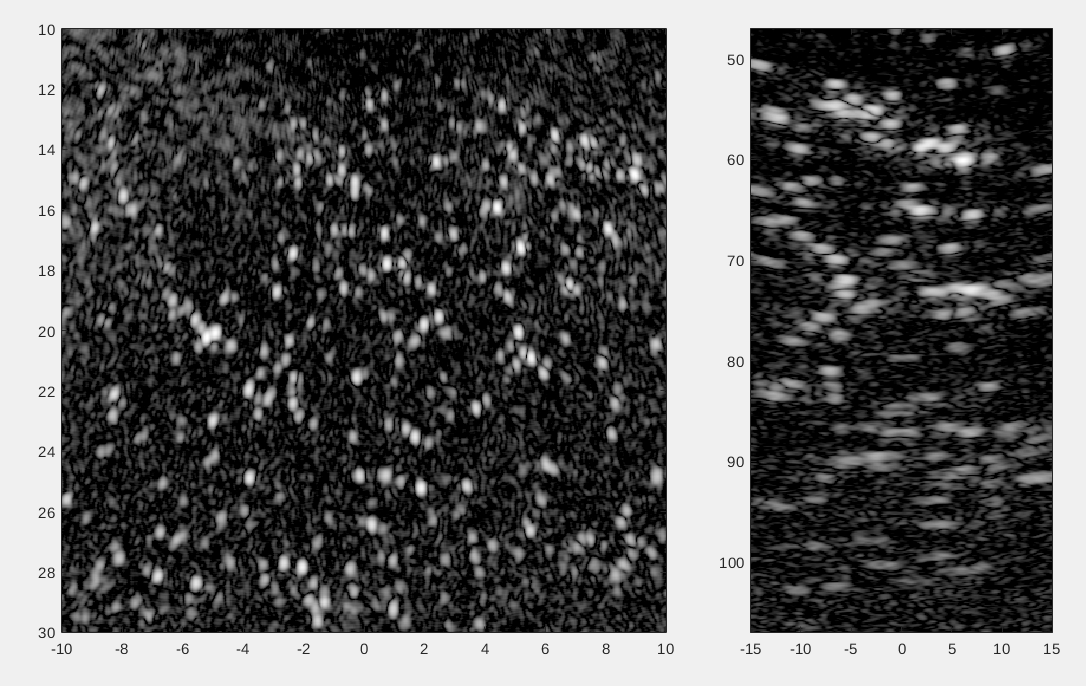}
      \caption{Generated B-Mode frames. \textbf{A:} Frame N149 of the \gls{hf} video simulation. \textbf{B:} Frame N4 of the \gls{lf} video simulation}
  \label{fig:bmode}
\end{figure}

Two randomly generated networks of \gls{mb} seeded vessel trees (\autoref{fig:net_struct}) were simulated using \gls{bff} as shown in Figure \ref{fig:net_struct}. The final networks were generated from multiple smaller networks to make localization and tracking difficult. No apparent structure within the networks is recognisable. \autoref{fig:bmode} shows the B-Mode images of both networks. It demonstrates the effect of adding \textit{Additive Colored Noise} and \textit{Gaussian White Noise} to the beamformed radio frequency signals. Through a \textit{Time Gain Compensation} the signal to noise ratio decreases with depth. The signal to noise ratio is higher in the low frequency dataset (\autoref{fig:bmode}, B) compared to the higher frequency dataset (\autoref{fig:bmode}, A).

\begin{figure}[H]
  \centering
    \includegraphics[width=0.8\columnwidth]{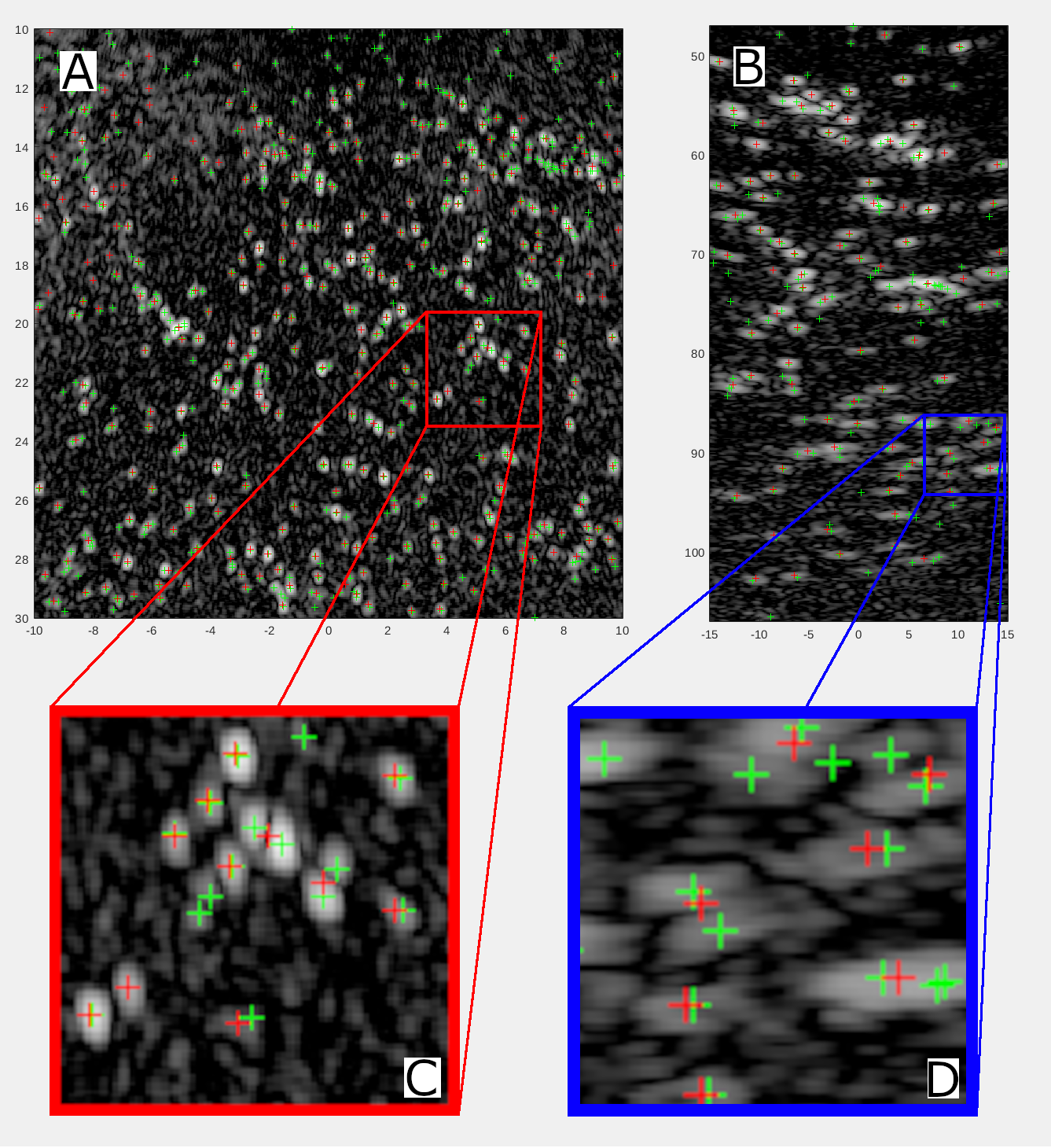}
      \caption{Superposition of a B-Mode frame and the bubble localizations. The ground truth is shown as a green cross, and the algorithm output is shown as a red cross. \textbf{A \& C:} \gls{hf} video frame N149 and detail of a region. \textbf{B \& D:} \gls{lf} video frame N4 and detail of a region.}
  \label{fig:localizations}
\end{figure}

\autoref{fig:localizations} shows the B-Mode frames, superimposed with both the ground truth locations and the predicted locations. The two expanded regions show in detail how accurate the predictions are. The prescence of \gls{fn} (isolated green cross) locations can be appreciated in both the \gls{hf} and \gls{lf} images. Additionally, both images (C \& D) show two neighbouring bubbles being identified as one.

\begin{figure}[H]
  \centering
    \includegraphics[width=1\columnwidth]{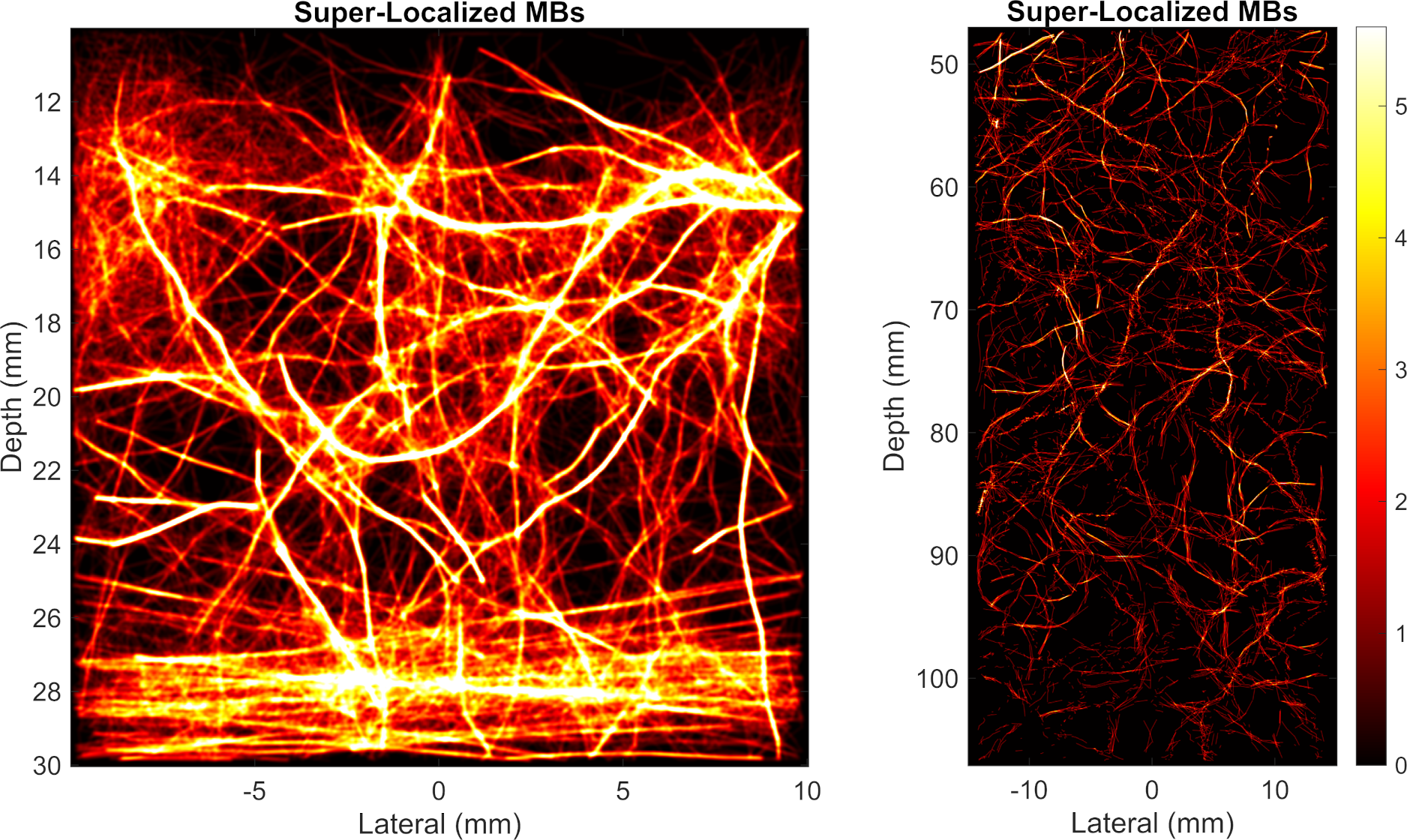}
      \caption{Resulting SR Images obtained from localization and tracking. \textbf{A:} \gls{hf} network. \textbf{B:} \gls{lf} network}
  \label{fig:sr_image}
\end{figure}

\begin{figure}[H]
  \centering
    \includegraphics[width=1\columnwidth]{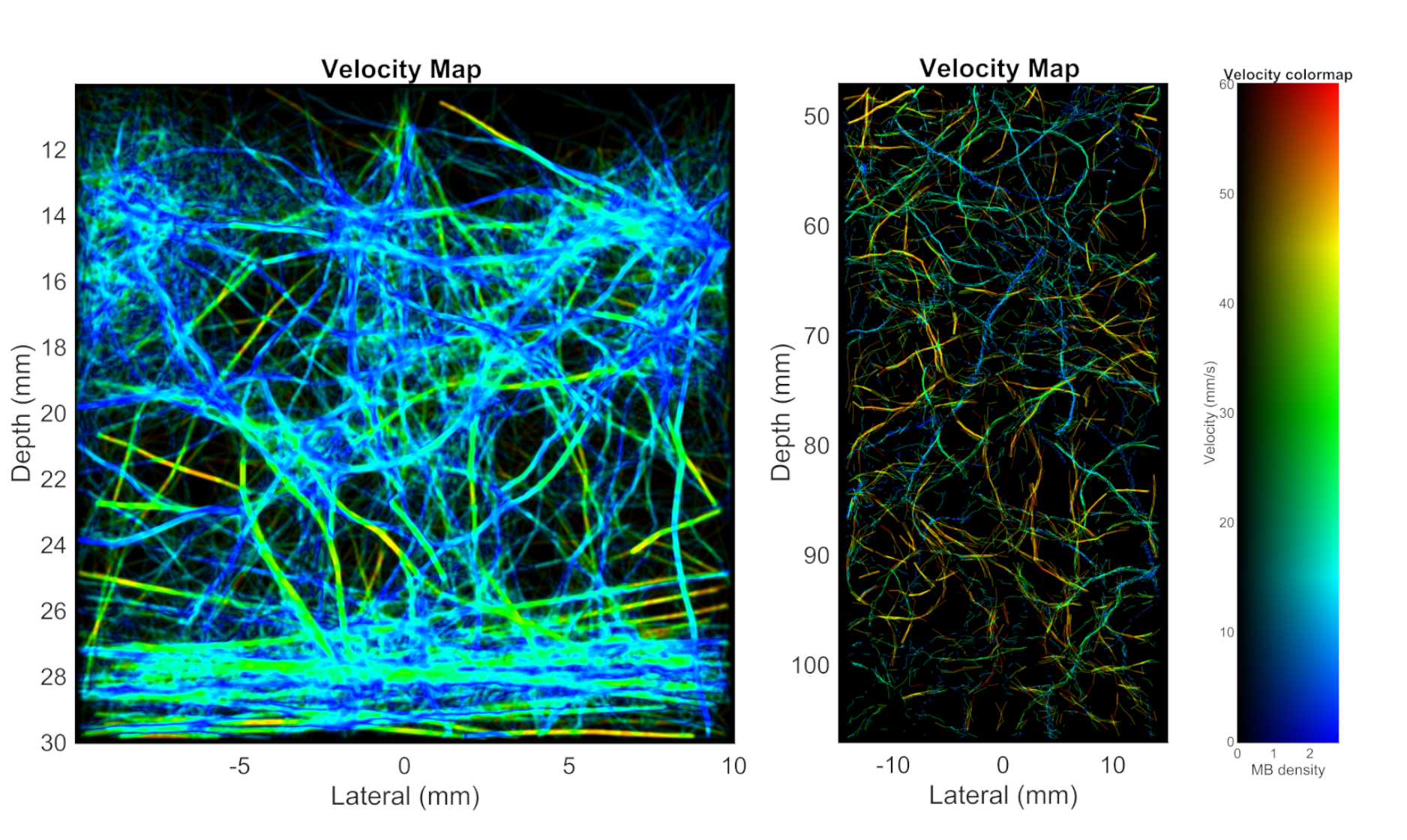}
      \caption{Resulting Velocity Map Image obtained from tracking. \textbf{A:} \gls{hf} network. \textbf{B:} \gls{lf} network}
  \label{fig:vel_image}
\end{figure}

The final superresolution image and its velocity map are shown in \autoref{fig:sr_image} and \autoref{fig:vel_image}. While the observed network resembles the one presented in \autoref{fig:net_struct}, the density of tracks is not the same, and large black gaps can be seen. This result is expected when the localization and tracking results (\autoref{tbl:eval_resutlts}) are taken into consideration. For the \gls{hf} dataset, a precision$=0.51$ indicates that around 50\% of the true localizations was not captured, and of the possible pairings to be made with the correctly localized positions, only 43\% were actually paired. On the other hand, the results also show a high tracking precision of $0.99$ for the \gls{hf} dataset, which means that only 1\% of the tracks present in the final image will be inexisting lines created by the algorithm.
The performance of the algorithm is worse for the \gls{lf} dataset despite the higher signal to noise ratio. This was expected because of its much larger psf. 

\begin{table}[ht]
    \begin{subtable}[h]{0.45\columnwidth}
        \begin{center}
            \begin{tabular}{ |l|c|}
                \hline
                    \multicolumn{2}{|c|}{\textbf{Localization}}
                \\\hline
                    \textbf{Precision} & 0.48
                \\\hline
                    \textbf{Recall} & 0.51
                \\\hline
                    \textbf{RMSE} & $ 135.79 \mu m$
                \\\hline
                    \multicolumn{2}{|c|}{\textbf{Tracking}}
                \\\hline
                    \textbf{Precision} & 0.99
                \\\hline
                    \textbf{Recall} & 0.43
                \\\hline
                    \textbf{$J_{map}$} & -0.71
                \\\hline
            \end{tabular}
            \caption{Low Frequency} 
            \label{tbl:eval_resutlts_LF}
        \end{center}
    \end{subtable}
    \hfill
    \begin{subtable}[h]{0.45\columnwidth}
        \begin{center}
            \begin{tabular}{ |l|c|}
                \hline
                    \multicolumn{2}{|c|}{\textbf{Localization}}
                \\\hline
                    \textbf{Precision} & 0.74
                \\\hline
                    \textbf{Recall} & 0.66
                \\\hline
                    \textbf{RMSE} & $ 47.34 \mu m$
                \\\hline
                    \multicolumn{2}{|c|}{\textbf{Tracking}}
                \\\hline
                    \textbf{Precision} & 0.56
                \\\hline
                    \textbf{Recall} & 0.44
                \\\hline
                    \textbf{$J_{map}$} & -0.57
                \\\hline
            \end{tabular}
            \caption{High Frequency} 
            \label{tbl:eval_resutlts_HF}
        \end{center}
    \end{subtable}
    \caption{Evaluation results} 
    \label{tbl:eval_resutlts}
\end{table}

\section{Discussion}

We developed \gls{bff}, a fully comprehensive simulation platform for \gls{ulm} algorithm development and evaluation. The framework consists of four key components: a microvascular structure generator, a flow simulator, an acoustic field simulator coupled with nonlinear microbubble dynamics and an evaluation pipeline for binary and quantitative assessment. The framework code is object oriented which makes it modular and easy to extend. \gls{bff} allows the user to create short, concise and powerful scripts that simulate dynamic \glspl{mb} excited by common ultrasound transducers and is open source.

The microvessel structure generator can be customized to yield networks that resemble realistic vascular structures, \autoref{fig:custom_nets}.A shows how parameters like the recursion level, can generate models with increasing vessel densities. Organic structures like the vessels structures seen on kidneys, can be replicated. \autoref{fig:custom_nets}.B shows how the ramifications of interlobar arteries of the rabbit's kidney can mimicked. To this point, only binary trees \autoref{fig:custom_nets} are capable of being generated, but the implementation can be extended to generate other networks. The addition of loop connections could help mimic intussusceptive angiogenesis for even more realistic structures. 


The incorporation of non-linear simulation of \glspl{mb} via Modified Rayleigh–Plesset \glspl{ode} has proven to generate realistic \glspl{psf} (\autoref{fig:psf_compare}) and enable simulation of coded transmission techniques such as \gls{am}. The simulation not only is realistic in terms of the shapes and sizes of the resulting \glspl{psf}, but also realistic \gls{rf} data is generated. This means that the datasets generated using \gls{bff} can be used to assess algorithms and techniques that require that type of input, for example beamforming. 

\gls{bff} helps address one important problem present in \gls{ulm} algorithm creation, evaluation and validation: the creation of realistic microvascular phantoms and data with ground truth. The realism of \gls{bff}’s simulations greatly facilitate the evaluation of \gls{ulm} algorithms, and the datasets generated by \gls{bff} framework has been used by the \gls{ultra-sr} challenge (\url{https://ultra-sr.com}) at the \gls{ieee} \gls{ius} 2022 to evaluate a number of localisation and tracking algorithms.

Even though this work focuses on \gls{ulm} as a proof of concept, \gls{bff} is not limited to this application only. Any contrast imaging related application can greatly benefit from datasets with ground truth for evaluation. Additionally, the accurate flow simulation is of special interest for applications like Ultrasound Vector Flow Imaging or Perfusion Imaging.


Data driven algorithms such as deep learning methods will benefit from this framework thanks to its compute time which is adequate for large dataset generation. The most time consuming step in the simulation framework is the acoustic field simulation. For the purpose of keeping it as open-source \gls{bff} uses the free version of FieldII\cite{jensen_field_1996}, further speed up can be achieved by using the commercial version. It should be noted that Field II does not simulate non-linear propagation of ultrasound. However, given the low mechanical index used for \gls{mb} imaging, the effect of nonlinear propagation, particularly when bubbles are excited at resonance frequency, is low \cite{tang_effects_2010}.

\section{Conclusion}

This work introduces \gls{bff}, a simulation framework for algorithm development and evaluation, with a focus on fast and large dataset creation, which will simplify the development and assesment of deep learning models and traditional algorithms for all the stages in the \gls{ulm} pipeline.
As a proof of concept, \gls{bff} was used for the ULTRA-SR competition at \gls{ieee} \gls{ius} 2022, proving that it can create random organic networks that resemble physiological structures seen in-vivo, it can yield flow simulations that are accurate representation so of actual microvascular flow, it uses simulation of \gls{mb} dynamics to generate realistic \glspl{psf} that resemble the ones seen on experiments, and it is suitable for large dataset generation thanks to its speed.

\section{Acknowledgements}

This work was supported by UK \gls{epsrc}.

The complete simulated dataset for \gls{ultra-sr} can be found on \url{https://doi.org/10.5281/zenodo.7271766}

\printbibliography

\end{document}